# Exciton-polariton condensation in a natural two dimensional trap

D. Sanvitto, A. Amo and L. Viña
*Dept. Física de Materiales. Universidad Autonóma. E28049 Madrid. SPAIN*

R. André
*CEA-CNRS. Institut NEEL-CNRS, BP166. 38042 Grenoble Cedex 9, FRANCE*

D. Solnyshkov and G. Malpuech
*LASMEA, CNRS and University Blaise Pascal, 24 Avenue des Landais, 63177 Aubiere Cedex, FRANCE*

Bose Einstein condensation of exciton-polaritons has recently been reported in homogeneous structures only affected by random in-plane fluctuations. We have taken advantage of the ubiquitous defects in semiconductor microcavities to reveal the spontaneous dynamical condensation of polaritons in the quantised levels of a trap. We observe condensation in several quantized states, taking their snapshots in real and reciprocal space. We show also the effect of particle interactions for high occupation numbers, revealed by a change in the confined wave function toward the Thomas Fermi profile.

PACS Numbers: 03.75.Nt; 71.36.+c; 78.47.jc; 71.35.Lk

## I. INTRODUCTION

Microcavity polaritons have been observed experimentally in 1992.[1] These quasi-particles arise from the strong coupling between the optical mode of a Fabry-Perot resonator and the excitonic resonance of a semiconductor quantum well. They behave at low density as bosonic particles, retaining from their excitonic component an interacting behaviour and from their photonic component a light effective mass and a lifetime in the picosecond range. A unique feature of polaritons is that both their life time and relaxation kinetics can be easily modified by changing the photon-exciton detuning, giving rise to different regimes of polariton condensation. Quasi-equilibrium Bose-Einstein condensation can be achieved under positive detuning in the *cw* pumping regime.[2-4] On the other hand the formation of a metastable condensate can be observed in the negative detuning case.[4,5] Here we use a CdTe microcavity, which is characterized by a strong exciton-photon coupling but also by strong in-plane in-homogeneities of the polariton energies which can be related to the fluctuations of the cavity mode energies.[2] These natural fluctuations influence the condensate wave function (WF) in real space,[2,6] which favours the formation of a glassy phase instead of a polariton superfluid.[7] Three dimensional optical confinement of a CdTe microcavity has been studied in micro-pillars under weak coupling,[8,9] showing a size-dependent energy splitting between photonic modes, and the existence of strong coupling has been demonstrated in pillars of II-VI[10] and III-V[11] microcavities and in traps produced by etching shallow mesas on top of the cavity layers,[12,13] and in quantum boxes.[14] However no observation of a condensate in a confined solid state system has been reported.

## II. EXPERIMENTAL DETAILS

In the experiments, polaritons are created in a CdTe-based semiconductor microcavity grown by molecular beam epitaxy. The sample consists of a $Cd_{0.4}Mg_{0.6}Te$ $2\lambda$-cavity with top (bottom) distributed Bragg reflectors of 17.5 (23) pairs of alternating $\lambda/4$-thick layers of $Cd_{0.4}Mg_{0.6}Te$ and $Cd_{0.75}Mn_{0.25}Te$. Four CdTe QWs of 50 Å thickness, separated by a 60 Å barrier of $Cd_{0.4}Mg_{0.6}Te$, are placed in each of the antinodes of the electromagnetic field. Exciton-photon coupling is achieved with a Rabi splitting of 23 meV.

The experiments are performed with a $Ti:Al_2O_3$ pulsed-laser of ~2 ps width and at energies well above the upper polariton dispersion, so that any memory of the laser coherence is lost. The sample is kept at 4 K and condensation is obtained around zero detuning between the exciton and the bare cavity dispersion. The photoluminescence is collected by an aspheric lens (objective) with high numerical aperture (NA=0.6) to allow detection of a wide range of k vectors (namely from -1.8 to +1.8 $\mu m^{-1}$). The polariton emission is collected with a second lens (ocular) into a 0.5 m spectrometer and analysed either through a high definition CCD or a streak camera for time-integrated or time-resolved images respectively. In the latter case the time resolution is about 7 ps, which allows the observation of the formation and decay of the polariton population. Images in reciprocal space are obtained positioning an additional lens, between the objective and ocular, with its focus at the Fourier plane of the objective lens.

We select a natural trap in the microcavity with an approximate size of 6 μm and deepness of 2 meV (consistent with a thickness fluctuation of the cavity by ~1 monolayer) and we perform off resonance pulsed excitation, with an excitation spot of 200 $\mu m^2$ focalised on the potential trap, obtaining real- and reciprocal (*k*)-space images of the polaritons' light emission, which reveal the energies and WFs of polariton condensation in the 6 first quantised eigenstates of the trap.

### III. RESULTS AND DISCUSSION

Figure 1a shows a time-integrated image of the polariton dispersion, at a slight negative detuning, below the condensation threshold, for which polaritons accumulate at the bottom of their parabolic dispersion. Note that the trap is not revealed by the emission, which is homogeneous inside the excitation spot in real space (Fig. 1b), as a result of the averaging of the different emission energies. Furthermore, the polariton linewidths (~1 meV) are dominated by the radiative decay $\Gamma_0$ and the phonon dephasing so that the quantized confined eigenstates cannot be resolved. Time-resolved images of the emission from the lower branch above the condensation threshold (1.1 mW), shown in Figs. 2a and 2b, allow the study of the condensation dynamics and reveal emission from 4 sharp lines. At short times (Fig 2a), a strong blue shift is obtained as a consequence of Coulomb interactions, mainly in the exciton reservoir, which monotonically decreases as the carrier density decays with time; however polaritons remain under strong coupling at all times as inferred by the lack of any saturation in the emission energies, which are also well below the bare photon energy (1.682 eV) excluding the possibility to observe simple lasing from the cavity optical field. The lowest energy states are mainly populated by stimulated scattering and become coherent soon after the pulse has arrived (see the time evolution of the linewidth for the ground state and the first excited state shown in Fig. 2d). Their coherence time is inversely proportional to their linewidth, which strongly drops above threshold, given by $\Gamma = \Gamma_0 / N_0 + g\sqrt{N}$,[15, 16] where $N_0$ is the population of a given state, $N$ is the total polariton population and $g$ the matrix element of interaction between polaritons,[17] which is about 0.1 µeV for the surface we study. From the jump of the intensity of emission, we estimate the population of the trapped states to be 40. This value is large enough to induce a strong narrowing of the emission lines, while the interaction energy between the condensed particles, $gN_0$, and the dephasing $g\sqrt{N}$, induced mainly by the polariton reservoir, remain small. Therefore a large reduction of the linewidth ensues, which allows the resolution of several confined states in the trap. The fact that not just the lower one, but several sharp emission lines are visible demonstrates the formation of various non-equilibrium confined condensates. The formation of such condensates is a unique feature of polaritons. It takes place when the relaxation kinetics of the particles is comparable with their lifetime.[4]

The use of a pulsed excitation from a mode-locked laser and a time-resolved detection set-up are of paramount importance for the direct observation of the different condensate states. On one hand it has been proven that intensity fluctuations in the excitation source result in particle number fluctuations in the exciton reservoir, which induce an additional inhomogeneous broadening of the condensed states.[18] In this sense pulsed mode locked lasers, as the one we use in this experiment, have reduced amplitude fluctuations (as low as single mode lasers), thus minimizing this effect.[19] On the other hand, even under conditions of stabilized intensity pumping, under pulsed excitation the blueshift induced by the reservoir population on the energy of the condensates changes in time, as discussed above (see Fig. 2a). A time resolved detection set-up enables us to register real and momentum space snapshots of the states at a given time delay, allowing us access to a higher energy resolution, which would be washed out by the continuous redshift of the emission if a time integrated scheme would have been used.

In the trap, assumed to be square for the sake of simplicity, the quantisation energy for this potential is approximately given by $E_0 = \frac{\pi^2 \hbar^2}{2mL^2} = 0.2$ meV ($m$ is the polariton mass, ~ 5 $10^{-5}$ of the free electron mass), which is much smaller than the potential depth, allowing for the formation of several confined discrete states in the trap. In an infinitely deep square well, the WFs and eigenenergies are given by $\psi_{n_x,n_y}(x,y) = \frac{2}{L}\sin\left(\frac{\pi n_x x}{L}\right)\sin\left(\frac{\pi n_y y}{L}\right)$ and $E_{n_x,n_y} = (n_x^2 + n_y^2)E_0$, respectively, where $n_{x,y}$ are positive integer quantum numbers. In a well of U = 2 meV only the first 6 states are confined with the following order of increasing energy: $\Psi_{1,1}$, ($\Psi_{1,2}$, $\Psi_{2,1}$), $\Psi_{2,2}$, and ($\Psi_{1,3}$, $\Psi_{3,1}$). Note that due to degeneracy [E($\Psi_{n_x,n_y}$)=E($\Psi_{n_y,n_x}$)] only 4 different energies are expected to be observable if the structure does not deviate significantly from the square well.

The balance between thermodynamics and kinetics determines the state's occupation at a given time. Figure 2c shows the ratio of the k=0 emission from state $\Psi_{2,1}$ over that of the state $\Psi_{1,1}$. Assuming that both levels are fed from the same reservoir, we can see that the relative gain of the stimulated scattering to each condensed state evolves in time, being more than three times larger right after the arrival of the excitation pulse and close to 1 at longer times. Even though state $\Psi_{2,1}$, due to its unresolved degeneracy, always presents a larger occupation than state $\Psi_{1,1}$, the linewidth of both levels shows a similar time evolution, as depicted in Fig. 2d. Moreover, the initial drop in the linewidth by a factor of 2.5 is a clear signature of spontaneous condensation, which takes place at the same time for both states.

To study the real- and *k*-space structure of the first 4 states, we take snap shots of the emission at a given time after excitation. The dispersion along $k_x$ at 47 ps, depicted in Fig. 2b, shows the same 4 states of Fig. 2a. The states are formed from the parabolic dispersion of the bare polaritons and appear as flat lines, a typical signature of localization. The fact that the intensity of the 2nd (4th) confined state is larger than that of the 1st (3rd) one, due to its double degeneracy, indicates that the symmetry of the confinement is not too different from that of a square. This degeneracy is expected to be lifted by any small asymmetry, but the splitting cannot be resolved experimentally. The horizontal lines in Fig. 2b show the state's energies calculated in the infinite well approximation, which fit satisfactorily with the experimental values.

In order to precisely model the shape of the condensate WFs, we use the coupled Gross-Pitaevskii (GP) equation for excitons and Schrödinger equation for polaritons:

$$i\hbar\frac{\partial}{\partial t}\psi_X(\vec{r},t) = \left(-\frac{\hbar^2}{2m_X}\Delta + g|\psi_X(\vec{r},t)|^2\right)\psi_X(\vec{r},t)$$

$$i\hbar\frac{\partial}{\partial t}\psi_{ph}(\vec{r},t) = \left(-\frac{\hbar^2}{2m_{ph}}\Delta + U(\vec{r})\right)\psi_{ph}(\vec{r},t)$$

(1)

where $V$ is the Rabi splitting, $U(\vec{r})$ is the in-plane square potential having a lateral size of 6 µm and deepness 2 meV, g is the interaction constant, $\psi_X - m_X -$ and $\psi_{ph} - m_{ph} -$ are the exciton and photon WFs –masses–, respectively. One should note that in the low density regime, interactions are negligible and the GP equation reduces to the Schrödinger equation. This allows us to find the eigenfunctions numerically by solving the coupled stationary equations on a grid. The resulting WFs in real and *k* space are shown on Fig. 3I panels (a,c,e,g) and (b,d,f,h), respectively. They do not strongly deviate from the infinite well WFs, being only slightly extended in the barriers. The corresponding experimental images of these states are shown in Figs. 3II. There is a remarkably good agreement between the experimental *k*-space images and the 6 first quantized eigenmodes calculated with the square potential. The lowest line is a single peak both in real (Figs. 3Ia,IIb) and *k* space (Figs. 3Ib,IIa) where the width of 0.37 µm$^{-1}$ fits well with the one of a particle localised within a 6 µm potential trap. This line corresponds to the state $\Psi_{1,1}$ having a confinement energy of ~ $2E_0$. The next state shows 4 clear points in *k* space (Figs. 3Id,IIc), located on a circle of radius 0.68 µm$^{-1}$. These images correspond to the two first, almost degenerate, excited states $\Psi_{1,2}$ and $\Psi_{2,1}$. We plot the sum of the intensities of emission of the two states calculated theoretically, assuming that the splitting is sufficient to eliminate interferences but not large enough to be resolved experimentally. Their energy spacing from the state $\Psi_{1,1}$ is 0.55 meV, very close to the expected value ~$3E_0$ = 0.6 meV. The simulation in real space (Fig. 3Ic) shows 4 points, similarly to the experimental image (Fig. 3IId) although the fourth is not visible. Figures 3I&IIe-f display the $\Psi_{2,2}$ WF where 4 points are again clearly visible in *k* space, located on a circle of 1.01 µm$^{-1}$. Its energy spacing with $\Psi_{1,1}$ is 1.3 meV, which is very close to the expected value of ~ $6E_0$ =1.2 meV. The last visible eigenstates are the two almost degenerate $\Psi_{1,3}$, $\Psi_{3,1}$. Now the *k* space images (Figs. 3Ih,IIg) show 5 peaks, a central one and 4 others located on a circle of 1.18 µm$^{-1}$. The energy spacing with $\Psi_{2,2}$ amounts to 0.36 meV, in agreement with the expected value ~ $2E_0$ = 0.4 meV. The agreement between theory and experiment is not only limited to qualitative resemblance of the images: the physical extents of the states also agree quantitatively, as shown by the black circles in *k* space. In spite of the satisfactory accordance, there are dissimilarities between calculated/experimental images, especially in real space (Figs. II). However, by adding a small perturbation to the symmetry of the potential in one of the corners, a good agreement also with the real space images can be obtained as shown in Figs. 3III. The perturbation affects stronger the real space images, while the *k* space ones (not shown) are not modified appreciably and the changes of the eigenenergies are also negligible.

These observations are unique from several points of view. The observed WFs are the collective WFs of condensed states. They keep their coherence against the numerous dephasing processes arising in solid state whereas the inter-condensate interactions remain negligibly small. Due to the non-equilibrium nature of the condensate, not only the ground state WF is observable but also those of higher energy states can be distinguished, as theoretically predicted in Ref. 20. This nature is also responsible for polariton lasing not coming from the ground state.[11] Another remarkable aspect is the very large spatial extension of the confined WFs, as a direct consequence of the very light effective mass of polaritons. For example, the WF of an electron in a quantum dot has an extension of a few nanometers, which renders impossible its direct observation. As to the atomic condensates, their excited states have been studied extensively,[21] however, there are no reports similar to the ones presented here.

We will now study how the condensates' WFs change when interactions between the condensate particles become important. In this high density regime, we expect the relaxation kinetics to be faster and the system to be partly described by its equilibrium features, as it was recently reported.[4, 5] In this framework, the use of the GP Eq. (1) is justified. The difference with respect to the case of Fig. 3 is that now the interaction term is playing a major role. The shape of the WF tends to minimize the repulsive interaction energy more than the kinetic energy. In the extreme case of large density, the kinetic energy becomes negligible and the WF takes the so-called Thomas-Fermi shape: the inverse of the potential profile, occupying all the available space in the potential well. On figures 4a,b we show experimental dispersions, above threshold, corresponding to a low (1.4 mW, 4a) and a large (4 mW, 4b) pump intensity. The time-integrated emission in the high pumping regime is blue shifted by ~0.2 meV with respect to the low pumping case. The intensity is increased by a factor of 50, resulting in a population of about 2000 within the trap, which is large enough to initiate a renormalisation of the shape of the WFs towards the Thomas-Fermi profile. Indeed, one can observe the shrinking of the emission in *k* space, for increasing pump power, which corresponds to an expansion in real space. The *k* space profiles of the WFs in both regimes are shown on Fig. 4c and compared with the theoretical ground state WF profiles, calculated by minimisation of the free energy [7] using two different values of particle density within the trap. The free energy of the polariton system can be written as

$$F = \int d\vec{r}\, \frac{\hbar^2}{2M} \Psi_0^*(\vec{r}) \Delta \Psi_0(\vec{r}) + \int d\vec{r}\, U(\vec{r}) |\Psi_0(\vec{r})|^2 + \frac{\alpha}{2} \int d\vec{r}\, |\Psi_0(\vec{r})|^4 \qquad (2)$$

Here we use the polariton basis considering only the lower polariton branch in order to simplify the numerical problem of minimization. $M$ is therefore the polariton mass and $\alpha$ the polariton-polariton interaction constant containing the excitonic fraction. The potential felt by the polaritons should, in principle, also be different from that of photons. However, we have checked that without interactions the eigenfunctions obtained diagonalizing the equations (1) or a similar equation for polaritons of the lower polariton branch are almost identical. This allows us to keep the potential $U(\vec{r})$ in the equation (2) the same as in the equation (1). The minimization of the wavefunction is performed numerically under the constraint of the constant particle number: $\int d\vec{r}\, |\Psi_0(\vec{r})|^2 = N$.

The *k* space narrowing is visible both in experiment and theory. The real space profiles of the calculated WFs are shown on Fig. 4d: the high density WF spreads in direct space in order to minimize the interaction energy between particles, approaching the Thomas-Fermi profile. This phenomena is in agreement with findings in atomic condensates,[22, 23] but was never reported for polariton condensates. In a homogeneous system, such a high density in the condensate would result in the formation of a superfluid state, which is not the case here because of the presence of the strong localising potential.

**IV. CONCLUSIONS**

In conclusion, we have studied experimentally and theoretically the spontaneous condensation of polaritons in a two dimensional potential trap. In the weakly interacting regime, different WFs of six condensates confined and quantized in the trap are observed in real and *k* space. This observation is made possible due to the coherence of the condensates and to a time-resolved analysis of the condensation dynamics. In the high excitation regime the WFs are dressed by interactions. The observed change of their shape toward a Thomas Fermi profile is well accounted for by the solution of the GP equation. Our experiments open new routes, not easily viable using atomic condensates, to tailor the spectral and spatial shape of macroscopic, coherent matter-waves.

ACKNOWLEDGMENTS

We thank C. Tejedor for a critical reading of the manuscript and P. R. Eastham for fruitful discussion. This work was partially supported by the Spanish MEC (MAT2008-01555 & QOIT-CSD2006-00019), the CAM (S-0505/ESP-0200), and the IMDEA-Nanociencia. D.S. thanks the Ramón y Cajal Program.

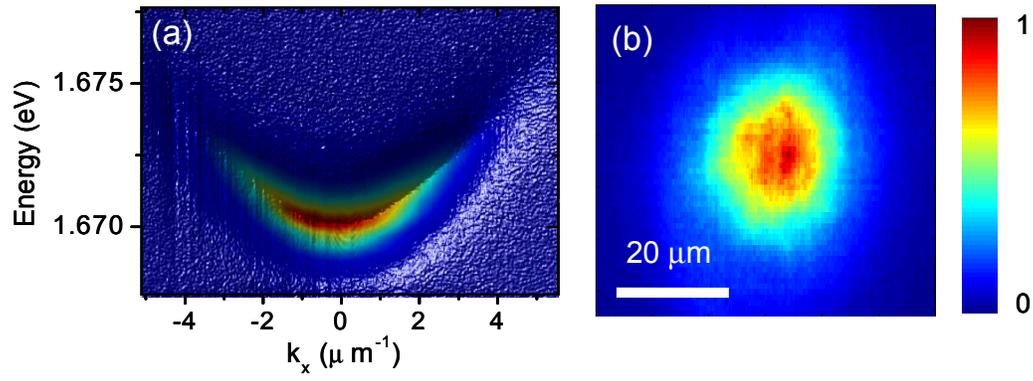

Figure 1. a) Time integrated polariton dispersion below the condensation threshold (1.1 mW) in a linear false colour scale. b) Real space image under the same conditions.

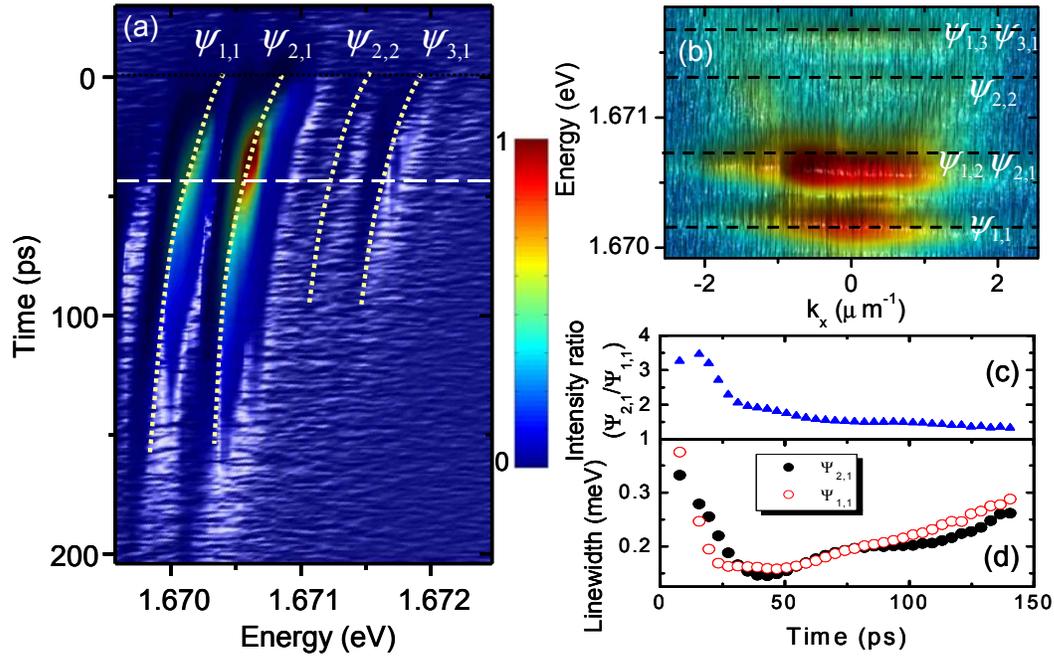

Figure 2. a) Time-evolution, after a pulsed excitation of 1.6 mW, of the condensates at k =0 in their confined states (dotted lines are guide to the eye). b) Dispersion of the condensate measured 47 ps after the pulsed excitation (dashed line in Fig. 2a). The states are labelled with their WFs. The false colour scale is logarithmic to reveal the highest confined states. The dashed lines show the energies of the WFs in an infinite quantum well. c) Time evolution of the intensity ratio, $\Psi_{2,1}/\Psi_{1,1}$. D) Time evolution of the linewidth of the states $\Psi_{2,1}$ (solid points) and $\Psi_{1,1}$ (open points).

.

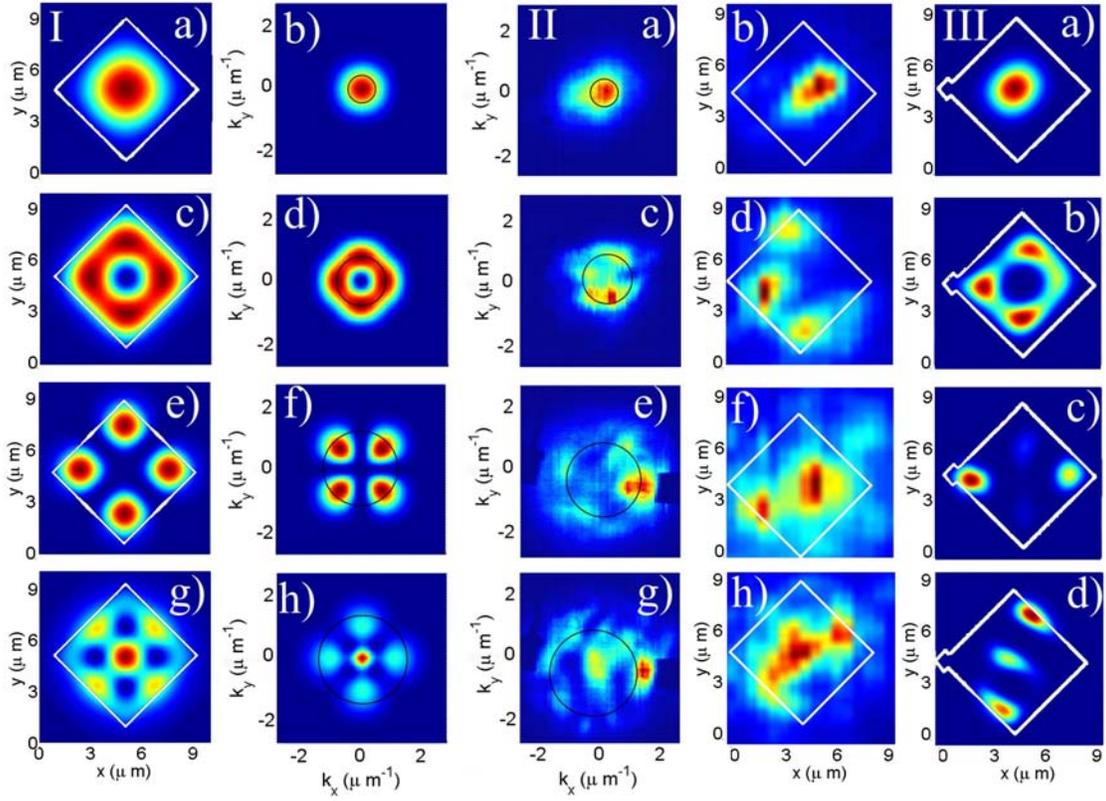

Figure 3 . Column I – simple theory, column II – experiment, column III – theory with perturbed potential. Ia,c,e,g), IIb,d,f,h), IIIa,b,c,d) - real space images of the WFs. Ib,d,f,h) and IIa,c,e,g) - *k* space images of the WFs. First row – $|\Psi_{1,1}|^2$, second row – $|\Psi_{1,2}|^2 + |\Psi_{2,1}|^2$, third row - $|\Psi_{2,2}|^2$, fourth row - $|\Psi_{1,3}|^2 + |\Psi_{3,1}|^2$. White solid lines: assumed potential boundary, black circles: physical extent of the states in the *k* space.

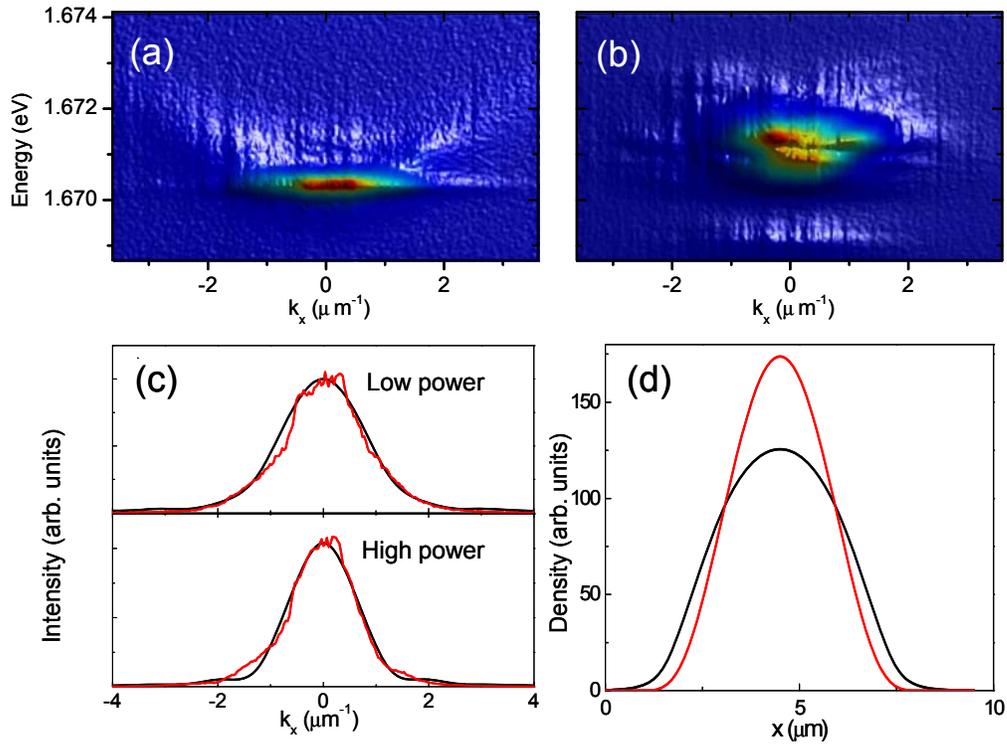

Fig. 4. Experimental dispersion images (a,b) and profiles in *k* (c) and real (d) space showing the variation of the condensate distribution with increasing density. Low density regime (1.4 mW)–panel a), top part of panel c), and the red curve on d)–. High density regime (4 mW)–panel b), bottom part of panel c), and the black curve on d)–. Panel c) shows the energy-integrated experimental profiles (red) and their theoretical counterparts (black), panel d) shows theoretical density profiles in real space.